\documentclass{sigchi-ext}

\copyrightinfo{Workshop paper presented at "Access InContext: Futuring Accessible Prototyping Tools and Methods", CHI'25, April 26, 2025, Yokohama, Japan. Submitted Feb 5, accepted Mar 1}

\usepackage[english]{babel}

\usepackage[T1]{fontenc}
\usepackage{textcomp}
\usepackage[scaled=.92]{helvet} 
\usepackage{graphicx} 
\usepackage{balance}  
\usepackage{booktabs} 
\usepackage{ccicons}  
\usepackage{ragged2e} 
\usepackage{dirtytalk} 
\usepackage[utf8]{inputenc}  
\usepackage{todonotes} 
\usepackage{float} 
\usepackage{subcaption} 
\usepackage{txfonts}

\usepackage{url}
\makeatletter
\g@addto@macro{\UrlBreaks}{\UrlOrds}
\makeatother

\usepackage{multirow} 
\usepackage{tabularx} 

\usepackage{color} 

\usepackage{enumitem} 

\usepackage[official]{eurosym}
\usepackage{siunitx} 

\usepackage[normalem]{ulem}

\usepackage{microtype}        
\usepackage[all]{hypcap}      
\usepackage{ccicons}          

\usepackage{cite} 

\title{Rethinking Accessible Prototyping Methods for Blind and Visually Impaired Passengers in Highly Automated Vehicles\\
\large \textit{Lessons from Two Participatory Design Workshops}}

\def\plaintitle{Rethinking Accessible Prototyping Methods for Blind and Visually Impaired Passengers in Highly Automated Vehicles\\
\large \textit{Lessons from Two Participatory Design Workshops}}

\def\plainauthor{Luca-Maxim Meinhardt, Enrico Rukzio}

\def\plainkeywords{highly automated vehicles, blind and visually impaired people, participatory design}

\title{\plaintitle}

\numberofauthors{2}
\author{
  \alignauthor{%
    \textbf{Luca-Maxim Meinhardt}\\
    \affaddr{Institute of Media Informatics} \\
    \affaddr{Ulm University, Germany} \\
    \email{luca.meinhardt@uni-ulm.de} } \vfil 
    \alignauthor{%
    \textbf{Enrico Rukzio}\\    
    \affaddr{Institute of Media Informatics} \\
    \affaddr{Ulm University, Germany} \\
    \email{enrico.rukzio@uni-ulm.de} }
    }

\definecolor{linkColor}{RGB}{6,125,233}
\hypersetup{%
  pdftitle={\plaintitle},
  pdfauthor={\plainauthor},
  pdfkeywords={\plainkeywords},
  bookmarksnumbered,
  pdfstartview={FitH},
  colorlinks,
  citecolor=black,
  filecolor=black,
  linkcolor=black,
  urlcolor=linkColor,
  breaklinks=true,
}

\begin{document}


\maketitle

\RaggedRight{} 

\begin{abstract}
Highly Automated Vehicles (HAVs) can improve mobility for blind and visually impaired people (BVIPs). However, designing non-visual interfaces that enable them to maintain situation awareness inside the vehicle is a challenge. This paper presents two of our participatory design workshops that explored what information BVIPs need in HAVs and what an interface that meets these needs might look like. Based on the participants' insights, we created final systems to improve their situation awareness. \\
The two workshops used different approaches: in the first, participants built their own low-fidelity prototypes; in the second, they evaluated and discussed the initial prototypes we provided. We will outline how each workshop was set up and share lessons learned about prototyping methods for BVIPs and how they could be improved.
\end{abstract}

\keywords{\plainkeywords}

\ccsdesc[500]{Human-centered computing~HCI theory, concepts and models}
\printccsdesc

\newpage
\section{Introduction}
Highly automated vehicles (HAVs) can improve the mobility independence of blind and visually impaired people (BVIPs) substantially~\cite{Kacperski.2024}. However, ensuring they gain trust and situation awareness inside the HAV is still an ongoing challenge~\cite{Brewer.2020}. In manually driven vehicles, BVIPs often rely on the sighted drivers to gain traffic information during the ride~\cite{Meinhardt.2024} or to 
gain situation awareness when exiting the vehicle~\cite{Brewer.2020, meinhardt_lightmyway_2025}. However, when HAVs become a reality, BVIPs will likely face situations alone without human assistance. Therefore, multiple researchers have already explored potential solutions to enhance BVIPs' situation awareness and trust:

For instance, Fink et al.~\cite{Fink.2023} studied mid-air haptics and tactile interfaces to improve BVIPs’ situation awareness while riding in HAVs. Other work has examined pre-journey mapping through smartphone applications~\cite{Fink_AUto_UI}, using vibrations on the phone to communicate route details. Additional studies have looked into vehicle localization solutions for ride-sharing scenarios, helping BVIPs locate and approach their vehicles independently~\cite{Fink.2023.3, RANJBAR2022100630}.
Regarding exiting the vehicle, the system ATLAS developed by Brinkley et al.~\cite{Brinkley.2019} utilizes computer vision to articulate the surroundings upon arrival at the destination~\cite{Brinkley.2019}. Most of these systems build on assumptions gained in earlier work that gathered and analyzed information needs of BVIPs~\cite{Bennett.2020, Brinkley.2020, Brinkley.2017}. Other research projects conducted their own exploratory user studies prior to implementing a solution (e.g.,~\cite{Fink.2023.2}).
In our research~\cite{meinhardt_lightmyway_2025, Meinhardt.2024}, we did similar by gathering qualitative data first and then built a final system based on the workshop's data.

This position paper presents insights from these two participatory design procedures that aimed to improve situation awareness for BVIPs inside HAVs as part of our research. Each procedure involved workshops with BVIPs to identify what information they needed, whether to gain an understanding of the traffic situation or to exit the vehicle safely. Based on the insights of these workshops, we then built prototypes that were evaluated in a follow-up user study.
We hope these lessons highlight how flexible, user-centered methods can help address the ongoing challenges in building accessible, effective solutions for BVIPs.

\section{Two Participatory Design Procedures}
This section briefly describes the participatory design processes in our two research projects, focusing on the qualitative data gathering during the workshops.

\subsection{Conveying Traffic Information \cite{Meinhardt.2024}}
\begin{figure}[h]
    \centering
    \small
    \includegraphics[width=0.92\linewidth]{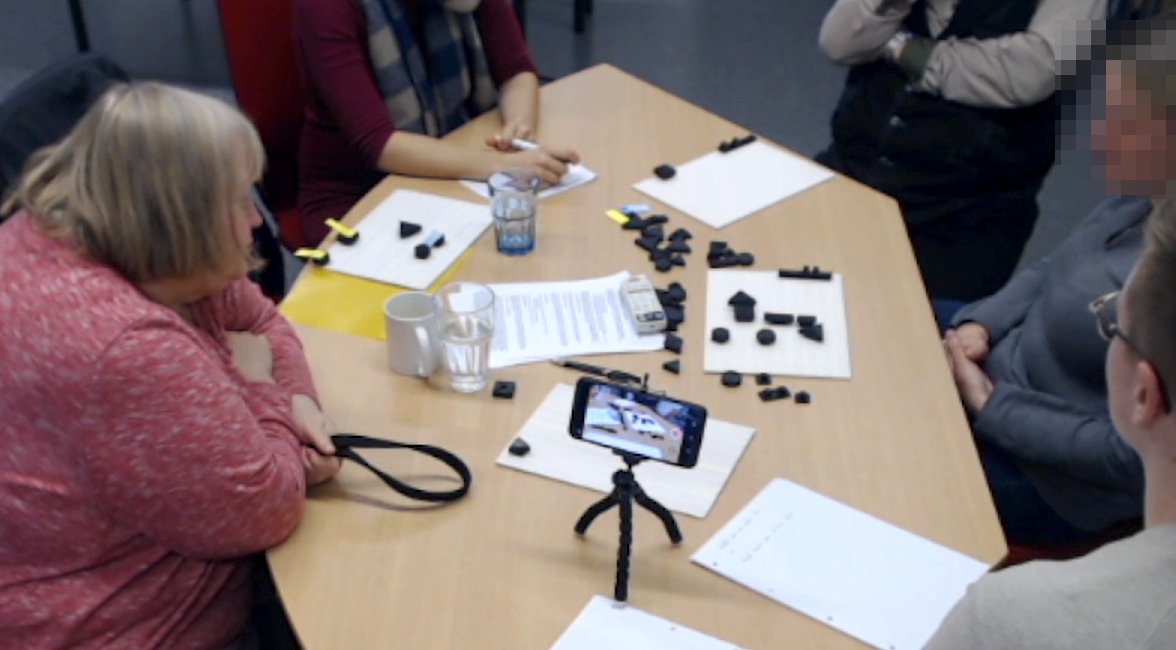}
    \caption{During the first project~\cite{Meinhardt.2024}, participants used 3D-printed elements to build and explain their own tactile interface designs that aim to convey traffic information.}
    \label{fig:workshop_1}
\end{figure}

In the first project, our goal was to convey traffic information through a tactile interface during HAV rides. We began by inviting N=4 BVIPs to a workshop to explore what information they considered most essential. After identifying their main information needs, we asked them to create a potential tactile interface using 3D-printed pieces that could be attached to boards covered with Velcro strips (see \autoref{fig:workshop_1}). 
Their input informed our final design of a servo-driven tactile display, \textsc{OnBoard}, which addressed the participants’ most relevant information needs and incorporated their interface ideas.
After the final prototype was built, we tested it against an auditory-only interface with N=14 BVIPs.

\subsection{Enhancing Situation Awareness When Exiting HAVs \cite{meinhardt_lightmyway_2025}}
\begin{figure}[h]
    \centering
    \small
    \includegraphics[width=0.92\linewidth]{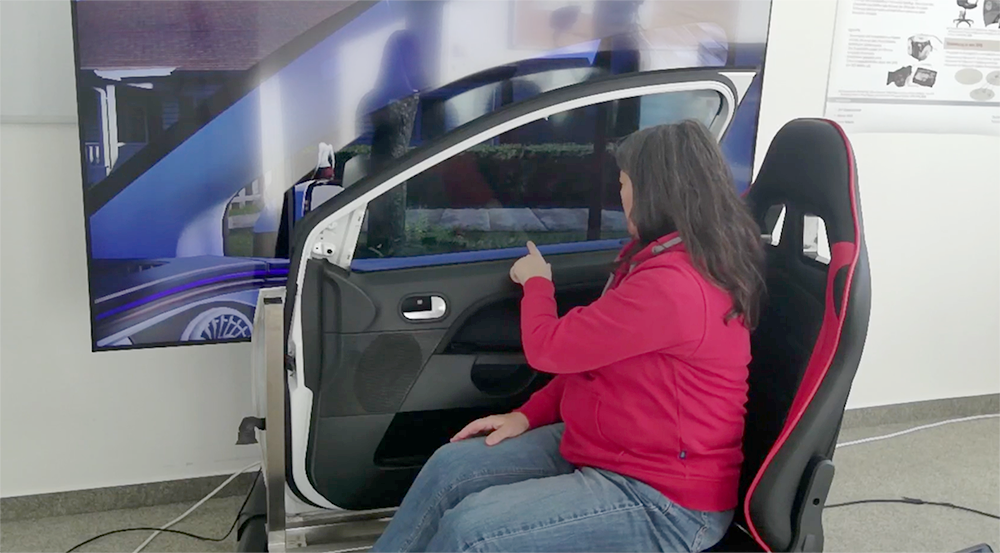}
    \caption{During the second project~\cite{meinhardt_lightmyway_2025}, participants examined our initial low-fidelity prototypes for exiting HAVs. In this figure, a participant is trying a touch interface on the vehicle's window.}
    \label{fig:workshop_2}
\end{figure}

The second project focused on helping BVIPs exit HAVs in unfamiliar locations. Unlike the first workshop, we presented three initial low-fidelity prototypes, each highlighting a different interaction approach to N=5 BVIPs, and collected qualitative feedback (see \autoref{fig:workshop_2}). 

After the individual sessions to explore each prototype, we gathered all participants again for a group discussion about their interface preferences. From these sessions, we learned which prototype features were most useful and combined them into a single multimodal system, \textsc{PathFinder}. This design choice was based on participants’ preferences for certain tactile elements, along with short audio cues for critical notifications.

Similar to the first project, we tested \textsc{PathFinder} with N=16 BVIPs against an auditory-only interface.

\section{Reflections on Accessible Prototyping}
Although both participatory design processes were similar, they differed in how we gathered the qualitative data. Below, we compare the two approaches.

\subsection{Hands-On Building vs. Ready-Made Prototype Testing}
In the first process~\cite{Meinhardt.2024}, participants described their information needs without being shown any fully functioning prototypes before. This gave us a broad sense of what they wanted, but it did not link their needs to specific interface solutions. Afterward, participants constructed their own interface ideas using 3D-printed elements, which helped them engage with the design task in a tangible way. However, it also limited them to the material that we provided, reducing their creativity as they might not have considered ideas outside the 3D-printed elements that we provided. In contrast, the second process introduced low-fidelity prototypes from the beginning. This led to more detailed feedback on each prototype’s strengths and weaknesses, as participants could immediately see how certain features behaved. However, it also offered fewer chances for participants to propose completely new concepts.

Our experiences suggest that tangible materials are valuable for eliciting detailed input from BVIPs~\cite{Holloway.2019, Holloway.2022}, but the nature and range of those materials should be carefully planned to avoid unintentionally limiting user creativity. A balance between open-ended materials (e.g., lego bricks) and more guided prototypes can help researchers capture both broad ideas and focused feedback.

One way to address BVIPs’ situation awareness needs in HAVs is to blend both design approaches: begin with a basic prototype and let participants modify or expand it on the spot, using items like 3D-printed pieces or clay that can be attached as needed. This gives participants a concrete starting point to explore but still allows them to shape the design based on their own ideas and preferences. Such an iterative process, however, requires multiple sessions and is thus more time-consuming. Yet, we believe that these repeated rounds of hands-on engagement yield richer insights and more targeted improvements, ultimately leading to prototypes that better match the diverse requirements of BVIPs.

\paragraph{Future Perspectives}
There is no need to reinvent the wheel for each new prototype. Nonetheless, many research projects produce functional systems that are no longer used once a study is completed, causing future work to revisit the same problems. While the building process of the prototype is often not the central part of the research, sharing these prototypes and the lessons learned from creating them can help others avoid repeated pitfalls. We therefore recommend providing open-source materials (such as code, 3D-print files, and hardware designs) so that future projects can build on existing work. We also encourage researchers to document their workshops and prototyping insights by capturing details about what failed, why it failed, and how it was fixed, giving others a chance to learn from these experiences and further strengthen accessible prototyping methods for BVIPs.

\section{Conclusion}
Designing accessible interfaces for BVIPs requires a balance between open-ended co-creation and guided prototyping. Our two participatory design methods show that early involvement of BVIPs yields deep insights but can limit creativity if materials are too constrained, while starting with ready-made examples encourages direct feedback but hinders broad exploration. A combined approach that offers flexible prototypes adapted through participant input may require multiple sessions, but promises more focused results. \\
However, many prototypes are discarded after one study, forcing future work to revisit the same challenges. Sharing these systems and workshop insights through open-source code, 3D print files, and detailed documentation helps others to build on existing progress rather than starting from scratch.

\bibliographystyle{SIGCHI-Reference-Format}
\bibliography{sample.bib}


\begin{thebibliography}{00}


\ifx \showCODEN    \undefined \def \showCODEN     #1{\unskip}     \fi
\ifx \showDOI      \undefined \def \showDOI       #1{{\tt DOI:}\penalty0{#1}\ } \fi
\ifx \showISBNx    \undefined \def \showISBNx     #1{\unskip}     \fi
\ifx \showISBNxiii \undefined \def \showISBNxiii  #1{\unskip}     \fi
\ifx \showISSN     \undefined \def \showISSN      #1{\unskip}     \fi
\ifx \showLCCN     \undefined \def \showLCCN      #1{\unskip}     \fi
\ifx \shownote     \undefined \def \shownote      #1{#1}          \fi
\ifx \showarticletitle \undefined \def \showarticletitle #1{#1}   \fi
\ifx \showURL      \undefined \def \showURL       #1{#1}          \fi

\bibitem{Bennett.2020}
{Roger Bennett}, {Rohini Vijaygopal}, {and} {Rita Kottasz}. 2020.
\newblock \showarticletitle{Willingness of people who are blind to accept autonomous vehicles: An empirical investigation}.
\newblock {\em Transportation Research Part F: Traffic Psychology and Behaviour\/}  {69} (2020), 13--27.
\newblock
\showISSN{13698478}
\showDOI{%
\url{http://dx.doi.org/10.1016/j.trf.2019.12.012}}


\bibitem{Brewer.2020}
{Robin Brewer} {and} {Nicole Ellison}. 2020.
\newblock {\em Supporting people with vision impairments in automated vehicles: Challenge and opportunities}.
\newblock {T}echnical {R}eport. University of Michigan, Ann Arbor, Transportation Research Institute.
\newblock
\showURL{%
\url{https://rosap.ntl.bts.gov/view/dot/56391}}


\bibitem{Brinkley.2020}
{Julian Brinkley}, {Earl~W. Huff}, {Briana Posadas}, {Julia Woodward}, {Shaundra~B. Daily}, {and} {Juan~E. Gilbert}. 2020.
\newblock \showarticletitle{Exploring the Needs, Preferences, and Concerns of Persons with Visual Impairments Regarding Autonomous Vehicles}.
\newblock {\em ACM Transactions on Accessible Computing\/} {13}, 1 (2020), 1--34.
\newblock
\showISSN{1936-7228}
\showDOI{%
\url{http://dx.doi.org/10.1145/3372280}}


\bibitem{Brinkley.2019}
{Julian Brinkley}, {Brianna Posadas}, {Imani Sherman}, {Shaundra~B. Daily}, {and} {Juan~E. Gilbert}. 2019.
\newblock \showarticletitle{An Open Road Evaluation of a Self-Driving Vehicle Human--Machine Interface Designed for Visually Impaired Users}.
\newblock {\em International Journal of Human--Computer Interaction\/} {35}, 11 (2019), 1018--1032.
\newblock
\showISSN{1044-7318}
\showDOI{%
\url{http://dx.doi.org/10.1080/10447318.2018.1561787}}


\bibitem{Brinkley.2017}
{Julian Brinkley}, {Brianna Posadas}, {Julia Woodward}, {and} {Juan~E. Gilbert}. 2017.
\newblock \showarticletitle{Opinions and Preferences of Blind and Low Vision Consumers Regarding Self-Driving Vehicles}. In {\em Proceedings of the 19th International ACM SIGACCESS Conference on Computers and Accessibility}, {Amy Hurst}, {Leah Findlater}, {and} {Meredith~Ringel Morris} (Eds.). ACM, New York, NY, USA, 290--299.
\newblock
\showISBNx{9781450349260}
\showDOI{%
\url{http://dx.doi.org/10.1145/3132525.3132532}}


\bibitem{Fink.2023.3}
{Paul~D.S. Fink}, {Stacy~A. Doore}, {Xue~(Shelley) Lin}, {Matthew Maring}, {Pu Zhao}, {Aubree Nygaard}, {Grant Beals}, {Richard~R. Corey}, {Raymond~J. Perry}, {Katherine Freund}, {Velin Dimitrov}, {and} {Nicholas~A. Giudice}. 2023c.
\newblock \showarticletitle{The Autonomous Vehicle Assistant (AVA): Emerging Technology Design Supporting Blind and Visually Impaired Travelers in Autonomous Transportation}.
\newblock {\em International Journal of Human-Computer Studies\/} (2023), 103125.
\newblock
\showISSN{1071-5819}
\showDOI{%
\url{http://dx.doi.org/https://doi.org/10.1016/j.ijhcs.2023.103125}}


\bibitem{Fink.2023}
{Paul D.~S. Fink}, {Anas {Abou Allaban}}, {Omoruyi~E. Atekha}, {Raymond~J. Perry}, {Emily~S. Sumner}, {Richard~R. Corey}, {Velin Dimitrov}, {and} {Nicholas~A. Giudice}. 2023a.
\newblock \showarticletitle{Expanded Situational Awareness Without Vision}. In {\em Proceedings of the 2023 ACM/IEEE International Conference on Human-Robot Interaction}, {Ginevra Castellano}, {Laurel Riek}, {Maya Cakmak}, {and} {Iolanda Leite} (Eds.). ACM, New York, NY, USA, 54--62.
\newblock
\showISBNx{9781450399647}
\showDOI{%
\url{http://dx.doi.org/10.1145/3568162.3576975}}


\bibitem{Fink.2023.2}
{Paul D.~S. Fink}, {Velin Dimitrov}, {Hiroshi Yasuda}, {Tiffany~L. Chen}, {Richard~R. Corey}, {Nicholas~A. Giudice}, {and} {Emily~S. Sumner}. 2023b.
\newblock \showarticletitle{Autonomous is Not Enough: Designing Multisensory Mid-Air Gestures for Vehicle Interactions Among People with Visual Impairments}. In {\em Proceedings of the 2023 CHI Conference on Human Factors in Computing Systems} {\em (CHI '23)}. Association for Computing Machinery, New York, NY, USA, Article 74, 13 pages.
\newblock
\showISBNx{9781450394215}
\showDOI{%
\url{http://dx.doi.org/10.1145/3544548.3580762}}


\bibitem{Fink_AUto_UI}
{Paul D.~S. Fink}, {H. Milne}, {A. Caccese}, {M. Alsamsam}, {J. Loranger}, {Mark Colley}, {and} {Nicholas~A Giudice}. 2024.
\newblock \showarticletitle{Accessible Maps for the Future of Inclusive Ridesharing}. In {\em 16th International Conference on Automotive User Interfaces and Interactive Vehicular Applications (AutomotiveUI '24)}. ACM, New York, NY, USA.
\newblock
\showDOI{%
\url{http://dx.doi.org/10.1145/3640792.3675736}}


\bibitem{Holloway.2022}
{Leona Holloway}, {Matthew Butler}, {and} {Kim Marriott}. 2022.
\newblock \showarticletitle{3D Printed Street Crossings: Supporting Orientation and Mobility Training with People who are Blind or have Low Vision}. In {\em CHI Conference on Human Factors in Computing Systems}, {Simone Barbosa}, {Cliff Lampe}, {Caroline Appert}, {David~A. Shamma}, {Steven Drucker}, {Julie Williamson}, {and} {Koji Yatani} (Eds.). ACM, New York, NY, USA, 1--16.
\newblock
\showISBNx{9781450391573}
\showDOI{%
\url{http://dx.doi.org/10.1145/3491102.3502072}}


\bibitem{Holloway.2019}
{Leona Holloway}, {Kim Marriott}, {Matthew Butler}, {and} {Samuel Reinders}. 2019.
\newblock \showarticletitle{3D Printed Maps and Icons for Inclusion}. In {\em The 21st International ACM SIGACCESS Conference on Computers and Accessibility}, {Jeffrey~P. Bigham}, {Shiri Azenkot}, {and} {Shaun~K. Kane} (Eds.). ACM, New York, NY, USA, 183--195.
\newblock
\showISBNx{9781450366762}
\showDOI{%
\url{http://dx.doi.org/10.1145/3308561.3353790}}


\bibitem{Kacperski.2024}
{Celina Kacperski}, {Florian Kutzner}, {and} {Tobias Vogel}. 2024.
\newblock \showarticletitle{Comparing autonomous vehicle acceptance of German residents with and without visual impairments}.
\newblock {\em Disability and rehabilitation. Assistive technology\/} (2024), 1--11.
\newblock
\showDOI{%
\url{http://dx.doi.org/10.1080/17483107.2024.2317930}}


\bibitem{Meinhardt.2024}
{Luca-Maxim Meinhardt}, {Maximilian R\"{u}ck}, {Julian Z\"{a}hnle}, {Maryam Elhaidary}, {Mark Colley}, {Michael Rietzler}, {and} {Enrico Rukzio}. 2024.
\newblock \showarticletitle{Hey, What's Going On? Conveying Traffic Information to People with Visual Impairments in Highly Automated Vehicles: Introducing OnBoard}. {\em Proc. ACM Interact. Mob. Wearable Ubiquitous Technol.\/} {8}, 2, Article 67 (may 2024), Article 67, 24 pages.
\newblock
\showDOI{%
\url{http://dx.doi.org/10.1145/3659618}}


\bibitem{meinhardt_lightmyway_2025}
{Luca-Maxim Meinhardt}, {Lina Wilke}, {Maryam Elhaidary}, {Julia von Abel}, {Paul Fink}, {Michael Rietzler}, {Mark Colley}, {and} {Enrico Rukzio}. 2025.
\newblock \showarticletitle{Light My Way: Developing and Exploring a Multimodal Interface to Assist People With Visual Impairments to Exit Highly Automated Vehicles}. In {\em Proceedings of the 2025 {CHI} {Conference} on {Human} {Factors} in {Computing} {Systems}}. ACM.
\newblock
\showISBNx{979-8-4007-1394-1/25/04}
\showDOI{%
\url{http://dx.doi.org/10.1145/3706598.3713454}}


\bibitem{RANJBAR2022100630}
{Parivash Ranjbar}, {Pournami~Krishnan Krishnakumari}, {Jonas Andersson}, {and} {Maria Klingegård}. 2022.
\newblock \showarticletitle{Vibrotactile guidance for trips with autonomous vehicles for persons with blindness, deafblindness, and deafness}.
\newblock {\em Transportation Research Interdisciplinary Perspectives\/}  {15} (2022), 100630.
\newblock
\showISSN{2590-1982}
\showDOI{%
\url{http://dx.doi.org/https://doi.org/10.1016/j.trip.2022.100630}}


\end{thebibliography}

\end{document}